\documentclass[9pt,a4paper]{extarticle}
\pdfoutput=1

\usepackage[T1]{fontenc}
\usepackage[utf8]{inputenc}   %% for UTF8-encoded accented characters in the Bibliography and elsewhere
\usepackage[pdftex]{graphicx}                        % if using pdftex
\usepackage[paper=a4paper, margin={1.75cm},centering]{geometry}
\usepackage{mathpazo}
\usepackage{inconsolata}                             %% a very nice monospace font -- replaces the tt font style
\usepackage{xcolor}
\usepackage{multicol}
\usepackage{hyperref}
\PassOptionsToPackage{hyphens}{url}

\usepackage{amsmath}
\usepackage{amsfonts}
\usepackage{amssymb}
\usepackage{amsbsy}
\usepackage{setspace}
\usepackage{xspace}                                %% for automatically applying an extra space (or not) after a character

\newcommand{\silent}[1]{}

\begin{document}

\begin{center}
   {\Huge\bf Graduate education in optics in Japan and the United States: impact of funding levels on educational structure} \\[8pt] Nathan Hagen \\ Utsunomiya University, Dept of Optical Engineering, Utsunomiya, Tochigi, Japan 321-8585 \\ email: \texttt{nh@hagenlab.org}
\end{center}

\vspace{15mm}
\setlength\columnsep{20pt}

\begin{abstract}
   {\noindent}We compare the optical science \& engineering graduate-level educational environments at two universities in two countries: Utsunomiya University in Japan, and the University of Arizona in the United States. Because the university education systems in the two countries are so different, we also explain how financial resources drive many of these differences and discuss how these impact student and faculty life.
\end{abstract}


\begin{multicols}{2}

\section{Introduction}

In the United States, centers for optical science such as the Institute of Optics at the University of Rochester, the Optical Sciences Center (OSC) at the University of Arizona, and CREOL at the University of Central Florida, have fulfilled a critical role in preparing skilled optical designers for the needs of technical industries, and for spawning successful startup companies. The success of these centers, both in fostering respected research and in educating skilled engineers, prompted many to realize that there is value in centering an educational curriculum around optical design and engineering, and not just spreading optics education piecemeal throughout the standard science and engineering curricula. 

In Japan, optics education has followed a very different path from that in the United States. Until recently, many famous names in the optics industry have been Japanese: Canon, Nikon, Olympus, Sony, Sigma, Tamron, Fujifilm, Mamiya, Konica, Minolta, Pentax, among many others. A side-effect of this success in industry, however, has been that optical science has played only a secondary role in Japanese universities --- typically as a subfield of physics, or of electrical engineering --- and optical design itself has played no part at all. Following the standard practice of post-war Japanese industry, Japanese companies hired Bachelor's degree (BS) or Master's degree (MS) graduates and put them through a multi-year training course in order to develop them into skilled optical designers. Companies thus preferred to foster expertise in-house and had little interest in supporting university optics education, since this would benefit potential rivals. As a result, it was not until 1993 that Tokushima University created Japan's first optics department, and not until 2008 that Utsunomiya University created an optics department providing students with an optical design curriculum.

One of the difficulties faced by anyone trying to form a new curriculum in optical design is getting optical design experts to switch from industry to academia. This is a problem faced by any new optics educational program anywhere in the world, but what has made this particularly difficult in Japan is that there is so little fluidity between academia and industry. Very few researchers are willing to switch from one to the other, making it difficult to find faculty able to teach optical design. Aggravating this problem are the basic facts that it is also notoriously difficult to publish research in optical design, and difficult as well to obtain competitive research grant funds for doing optical design work. While the Japanese Ministry of Science and Education, the largest source for university research funding in Japan, has an ``optical engineering'' research category, is in fact devoted to quantum optics research rather than optical design. This means that the research funding for a Japanese professor focused on optical design will come almost entirely from companies, who are less focused on publication as a research output, and who are primarily focused on practical applications.

In the discussion below, we survey the overall educational environment experienced by faculty and students at OSC in Arizona, and at the Center for Optics Research and Education (CORE) in Utsunomiya --- the two optical science centers having similar educational and research goals, but working in very different academic environments of Japan and the U.S. In the discussion below, we first look into the reasoning for why it is useful to create an optics department in the first place, before going on to survey the differences in financial resources between the two countries, and how these impact faculty, students, and the curriculum.

\section{Why should there be an independent department of optical engineering?}

For a long time, optics has been taught as a segment of the physics curriculum, or as an elective part of the electrical engineering curriculum. Why, then, should we need to teach it as an independent field of its own? This is an important question, and one which the Optical Sciences Center in Arizona faced many times in its early days.\cite{WyantPersonal} This is best answered by analogy, by asking the equivalent question of why we should need to teach, say, electrical engineering as an independent field. The answer to this analogous question suggests the answer for optical engineering as well.

Clearly, it is useful for many students of physics to learn how to design and build electrical circuits, and some of them do. This skill helps them to build better instruments, and to better understand the instruments that they are using in their own research projects. On average, however, these physics students will not be as adept at it as the typical electrical engineering student will be, since the latter is more narrowly focused on electrical circuit design or on subjects closely related to it. Students of physics, on the other hand, will therefore study circuit design as a secondary undertaking, and will likely be working on research that is only tangentially related to circuit design.

In a similar way, there are many physics students using optical instruments, and therefore who learn optical design principles in order to understand and build better instruments. While they can become adept at it, the tradeoff is that time spent learning optical design is time they cannot spend learning other subjects. Since learning to do optical design well requires years of study, this is not an easy tradeoff to make. These physics students would not be surrounded by colleagues pursuing similar studies, as they would if they were studying in an optical design curriculum. As a result, reaching the same level of skill in optical design would take them more time than if they could consult with teachers, colleagues, and friends at difficult points.

Clearly, for students of physics, electrical engineering and optical engineering are of a similar category --- useful but difficult fields to master if one is pursuing them as secondary subjects rather than primary ones. So, the common impression that optical engineering is not really an independent field is problem one of scale rather than of category --- the reaction that the field is too small to justify independence. The successes of OSC in Arizona, the Institute of Optics in Rochester, and CREOL in Florida, have demonstrated that this is not the case. Companies actively seek out graduates from these schools in order to take advantage of their understanding of optical systems and ability to do optical design.

\section{How funding shapes the educational environment}

Having established the need for an optical engineering curriculum, we can establish what form it should take. While the basic learning requirements for students are naturally very similar between the two countries, the learning experience of students in Japan differs significantly from those in the U.S.

These differences are driven, in large part, by funding levels, which play a large role in determining how academic departments are run. The financial resources passing through an academic department at a modern research university in the United States are roughly ten times those typically available to a comparable department in Japan. Most of these additional funds are directed towards financially supporting students during their period of study --- a practice that has become widespread in most of the industrialized countries, and is now regarded by many American graduate students in science and engineering as an entitlement. 

Because this student support plays a central role in modern science curricula, a concrete example helps show how it impacts the way that universities are run. For a PhD student at the University of Arizona in 2007, this financial support included a tuition waiver (${\sim}$\$20,000/year) and a stipend to cover basic expenses: ${\sim}$\$20,000/year, increasing to ${\sim}$\$22,000/year for students with an MS degree. Thus, a professor wishing to hire a PhD student to join his or her lab as a research assistant (RA) would have to pay those fees plus overhead costs --- a total of roughly \$50,000 per year.\cite{Heckel1995} Although this support is for only a single student, this is already more funding than a typical Japanese research lab receives in a year. A successful lab in the U.S. would have a cadre of at least five students, amounting to a minimum funding level of about \$250,000/year, without including equipment costs. While this is an extraordinary amount of money for a Japanese research lab, it is merely ordinary in the United States. (And readers should note that these numbers correspond to my period as a graduate student. Now, 15 years later, the costs have increased substantially.)

An American professor with insufficient funds can, with permission from the department, ask his students to work as teaching assistants (TAs), in which case the department covers much of the financial support, as long as there are sufficient TA positions available. For students, these TA positions are naturally less desirable than RA positions. 

%For the 2023--2024 school year, the tuition at UA for in-state resident MS students is \$14,000/year [in-state tuition+fees]. For out-of-state students, it is \$32,000/year!

Because of this expectation of support, the number of PhD students that an academic department allows to enroll each year is determined almost entirely by the amount of funding that the department's professors can provide for their support.\cite{Graddy-Reed2021} The financial support reduces the pressure that students feel to graduate as quickly as possible. While this frees them to work hard on learning, it can also be an incentive for them to be lazy. However, receiving a stipend as a professor's RA also means that any professor disappointed with a student's work can withdraw the stipend. Just as in a company, students can be fired. This helps to motivate students to work hard at their research.

In sharp contrast to students in the U.S., only few Japanese PhD students can expect to receive a monthly stipend, with these few deriving primarily from very competitive national and company-sponsored scholarships. Thus, students (or more likely their parents) are expected to pay their own way throughout their PhD studies. At Utsunomiya University, the typical annual tuition is {\yen}536,000/year (equivalent to US\$3480 as of the year 2025), and basic living expenses are about {\yen}110,000/month (US\$700/month). While these costs are considerably lower than they are for a typical university in the United States, they remain a severe financial barrier for students considering a research career.

The lack of basic financial support is one of the reasons why there are significantly fewer PhD students in Japan than there are in the United States. A second reason is that Japanese students pondering a research career in industry have long been told that companies prefer to educate their employees in-house rather than hire already-trained researchers. This is slowly changing, with the decline in corporate prosperity and the consequent reduction in spending for companies' internal education. But many students still feel that there is a stigma against PhD students trying to enter a research career in industry.\cite{NikkeiAsia}

One consequence of the excellent financial support provided by universities in the U.S.\ is that there is heavy pressure placed on professors to obtain research funding. Beyond the initial startup funding supplied to a new professor to get his or her research lab off the ground, universities provide very little direct research funding to professors. To compensate, researchers spend a large portion of their time --- significantly more of it than their peers in Japan do --- writing grant applications and searching for alternative sources of funding.

In contrast, professors in Japan generally receive modest annual research funding support directly from their universities. While this funding is not large --- {\yen}600,000 to {\yen}1,500,000 per year is common --- it is still enough to allow for basic maintenance of a lab. In addition, since students are almost always the biggest expense of a U.S.\ lab's funds, the fact that students in Japan are generally not funded through research grants means that a small research subsidy for a Japanese lab can go a long way. This subsidy also arrives annually without tedious grant applications. Finally, the certainty of receiving this basic level of funding allows Japanese professors a degree of freedom in choosing their research topics that an American professor might envy, since it allows them to focus on topics that are scientifically interesting rather than financially promising. While these two categories overlap, they are not at all the same thing.

%Another striking difference between Japan and the U.S. is that professors in Japan typically do not start their careers as fully independent researchers. A new assistant professor usually joins a senior professor's lab, and only becomes fully independent upon promotion to associate professor. Even more, the ability of a professor to mentor different levels of students is achieved in multiple small steps. At first, a new assistant professor can only advise BS students and not MS students. Upon achieving some experience as a mentor, an assistant professor can apply for an upgraded status which, if approved, allows one to be a secondary but not primary advisor for MS students. Becoming a primary advisor for MS students requires another application for upgraded status, which also confers the ability to be a secondary (but not primary) advisor to PhD students. One final upgraded status is then necessary in order to have one's own PhD student --- a status which is often restricted to tenured professors. One can expect each upgrade in status to take one or two years if time.

\section{Differences in coursework and learning experience}

Among many research universities in the U.S., almost every course with more than 10 students can expect to be assigned a student teaching assistant. Having TAs provides a means of financially supporting PhD students, while also freeing up a professors' time. Diligent students in the U.S.\ can expect to spend perhaps 4 hours per week, per class, on homework assignments. While many academics have argued against this system as a waste of students' time,\cite{Mazur1996} others have also argued for its importance in learning technical material.\cite{Byun2014} The standard lecture course in Japan also involves shorter in-class time: 1.5 hours of lecture time per week for 15 weeks --- nearly half that of the equivalent course in the U.S.

The issue of who does the grading reveals another contrast between the U.S.\ and Japan. Professors' limited time means that the few TA positions provided in Japan naturally leads to fewer and shorter homework assignments. Few professors are willing to spend hour after hour on grading homework assignments, week after week, so that problem-solving time spent by Japanese students is perhaps one-fifth that of students in the U.S. We can conclude that without the substantial support for TAs at U.S.\ universities, the demanding homework system there would likely evaporate.

A common feature of science and engineering departments in Japan is that fourth-year under\-grad\-uates are required to join a research laboratory and spend a year working on a senior thesis project. A consequence is that Japanese professors spend a substantial amount of their time advising these undergraduate students on their senior research projects. Thus, while Japanese professors spend less time lecturing, they spend more time mentoring, then their American counterparts. This senior year is often a Japanese student's first experience with self-directed learning, working with lab equipment, giving presentations, and writing a long report. For some, the newness and the sudden lack of immediate direction comes as quite a shock, and some students fail to progress with their projects. Whereas such students in the U.S.\ would simply drop out of the program, this is quite rare in Japan, and reflects a difference in education philosophy. As a result, Japanese professors work hard to help the lowest-performing students and go to extraordinary efforts to make sure that they obtain their degree. The aggressive student attrition pursued by many prestigious institutions in the U.S.\ is utterly absent in Japan.

The decision of a CORE student to continue from the MS program directly into the PhD program reflects how deeply they are interested in academic research, but also on their degree of financial independence. Most MS students graduate with plans to work at a company. At some later point, some decide that they want to pursue a PhD degree, but without leaving their current employment. As a result, a significant number of PhD students in Japan are actually employed full-time at companies and are enrolled part-time in their PhD studies. Naturally, due to their work schedules, these students tend to have difficulty directing their full attention to their research.

% For both MS and PhD students, the lack of a stipend to provide basic financial support influences their research motivation. For some students, there is a tendency to see this period of their university studies as a pause before their ``real'' lives begin as company employees. It is easy to take this too far and to blame the lack of financial support for students' lack of enthusiasm. I rather suspect that the problem is only partly one of financial incentive. Utsunomiya University is a regional school and not a nationally-renowned one, so that the types of students that come to us are generally not the top academically motivated. Of course, this is a matter of averages, and we have encountered our share of diligent students.

One conspicuous problem faced by graduate students in Japan is an abhorrence of writing research papers. The problem is certainly bad enough in the United States, and is reflected in the poor writing one finds in journal articles even by native speakers of the language. However, in Japan the reluctance towards writing reaches problematic levels. It is an unfortunate fact of life that non-English-speaking parts of the world must learn to write in a foreign language in order to make their scientific work known.

\section{Conclusion}

While it is easy for Japanese universities to look to their counterparts in the U.S.\ as examples to emulate, the difference in financial resources between the two means that some features of U.S.\ universities are impossible to replicate. Thus, it is instructive to see how finances shape what we now see (in the United States, and widely in Europe as well) to create the now-standard environment for modern science academies. Those who are exposed to only this one environment can easily miss the fact that there are alternative approaches that operate under more modest financial means. While most modern science educators measure success against the standards developed in the U.S.\ and Europe, and therefore see differences from the standard as flaws, the alternative approach is not all bad by comparison. Particularly in the U.S., the standard approach is great for supporting students, but tough on research freedom.

%\paragraph{Data Availability.} Data sharing is not applicable to this article, as no new data were created or analyzed. 

% %\bibliographystyle{/home/nh/mytemplate}
% \bibliographystyle{spiejour_bibulous}
% \bibliography{/home/nh/articles,./extras}    % bibliography file

\end{multicols}

\end{document}